\documentclass[journal,comsoc]{IEEEtran}
\usepackage{amsmath,amsfonts}
\usepackage{algorithmic}
\usepackage{algorithm}
\usepackage{array}
\usepackage[caption=false,font=normalsize,labelfont=sf,textfont=sf]{subfig}
\usepackage{textcomp}
\usepackage{stfloats}
\usepackage{url}
\usepackage{verbatim}
\usepackage{graphicx}
\usepackage{xcolor}
\usepackage{cite}

\usepackage[binary-units = true]{siunitx}
    \DeclareSIUnit{\belmilliwatt}{Bm}
    \DeclareSIUnit{\belisotropic}{Bi}

\begin{document}

\title{Little or no equalization is needed in energy-efficient sub-THz mobile access}

\author{Lorenzo Miretti, Thomas Kühne, Alper Schultze, Wilhelm Keusgen, Giuseppe Caire, Michael Peter, Sławomir~Sta\'nczak,  Taro Eichler 
\thanks{L.~Miretti, A.~Schultze, M.~Peter, and S.~Sta\'nczak are with the Fraunhofer Heinrich Hertz Institute (email: \{lorenzo.miretti, alper.schultze, michael.peter, slawomir.stanczak\}@hhi.fraunhofer.de). L.~Miretti and S.~Sta\'nczak are also with the Technische Universität Berlin. T.~Kühne, W.~Keusgen, and G.~Caire are with the Technische Universität Berlin (email: \{thomas.kuehne, wilhelm.keusgen, caire\}@tu-berlin.de). T.~Eichler is with Rohde \& Schwarz, Munich (email: taro.eichler@rohde-schwarz.com). The authors acknowledge the financial support by the Federal Ministry of Education and Research of Germany in the programme of “Souverän. Digital. Vernetzt.” Joint project 6G-RIC, project identification numbers: 16KISK020K, 16KISK030.}
}

\maketitle


\IEEEpubid{\begin{minipage}{\textwidth}\ \\[12pt] \centering
  This work has been submitted to the IEEE for possible publication.\\ Copyright may be transferred without notice, after which this version may no longer be accessible.
\end{minipage}}

\begin{abstract}
By trading coverage and hardware complexity for abundance of spectrum, sub-THz mobile access networks are expected to operate under highly directive and relatively spectrally inefficient transmission regimes, while still offering enormous capacity gains over current sub-6GHz alternatives. Building on this assumption, and supported by extensive indoor directional  channel measurements at \SI{160}{\giga\hertz}, this study advocates the use of very simple modulation and equalization techniques for sub-THz mobile access. Specifically, we demonstrate that, under the aforementioned transmission regimes, little or no equalization is needed for scoring significant capacity gain targets. In particular, we show that single-carrier or low-number-of-subcarriers modulations are very attractive competitors to the dramatically more complex and energy inefficient traditional multi-carrier designs.
\end{abstract}

\begin{IEEEkeywords}
sub-THz, energy efficiency, single-carrier, waveform design, 6G.
\end{IEEEkeywords}

\section{Energy-efficient sub-THz mobile access}
\IEEEpubidadjcol
\label{sec:intro}
\subsection{Spectral efficiency is neither needed nor welcome}
The exploitation of the large portions of available spectrum in the sub-THz band (90-300\,GHz) is one of the most promising directions for enhancing the capacity of current mobile access networks \cite{rappaport2019wireless,rajatheva2020scoring}. In contrast to current sub-6GHz networks, for which a 100-fold capacity increase can only be achieved by means of extreme spatial multiplexing and very complex modulation schemes, sub-THz networks can score this ambitious goal by transmitting fewer simultaneous data streams with relatively low spectral efficiency ($\approx$1-3 bit/s/Hz). These observations are best illustrated by focusing on the following approximate formula for the network capacity:
\begin{equation}
C =  B\cdot M \cdot \mathrm{SE} \quad \text{bit}/\text{s}/\text{km}^2,
\end{equation}
where $B$ is the signal bandwidth in Hz, $M$ is the number of spatially multiplexed streams per $\text{km}^2$, and $\text{SE}$ is the per-stream spectral efficiency in bit/s/Hz. In excellent conditions, current 5G sub-6GHz networks approximately support $B\approx 100$\,MHz, $M \approx 16$, and $\mathrm{SE}\approx 3-6$\,bit/s/Hz, by using massive MIMO arrays with $64$ antennas and multi-carrier $256$-QAM modulated signals. A 100-fold capacity increase at sub-6GHz frequencies would require a 100 times larger aggregate spectral efficiency (i.e., the product of $M$ and $\mathrm{SE}$), which is very challenging to achieve due to the problematic interaction between $M$ and $\mathrm{SE}$ caused by interference, and since $\mathrm{SE}\approx 10$\,bit/s/Hz is already close to its practical limit caused by hardware imperfections. In contrast, by moving to sub-THz frequencies, for which bandwidths up to $B\approx 30$\,GHz are conceivable, the same goal could be achieved with a dramatically more relaxed requirement on both $M$ and $\mathrm{SE}$.

Admittedly, if the target $100$-fold capacity increase must be realized in an energy efficient manner, finding the optimal trade-off between bandwidth and spectral efficiency is a highly non-trivial task. For instance, the best known analytic tools do not cover the energy consumed by the hardware \cite{verdu2002}. The present study focuses on sub-THz mobile access, i.e., on the exploration of the very large bandwidth extreme of this trade-off. In this regime, restricting to modulation schemes with low spectral efficiency is not only sufficient but also of paramount importance, since the resulting relaxation of the hardware requirements offers a unique opportunity for developing sub-THz transceivers with tolerable energy consumption \cite{rajatheva2020scoring}. 

\subsection{Highly directive steerable beamforming antennas}
\IEEEpubidadjcol
To guarantee reasonable coverage at such high frequencies and large bandwidths with acceptable  radiated power, highly directive antennas must be used \cite{rappaport2019wireless}. For instance, upgrading an ideal sub-6GHz free-space link through a $100$-fold increase in frequency and bandwidth, while keeping the same coverage,  radiated power, and target spectral efficiency, would require a directivity gain of about $60$\,dB. For this reason, even by considering significantly more relaxed coverage and spectral efficiency requirements than their sub-6GHz counterparts, sub-THz mobile access networks will likely need directive antennas ($\approx 10-30$\,dBi gain) at both the transmit and receive ends. Furthermore, due to user mobility, high directivity must be dynamically realized via steerable beamforming antennas. This poses a series of remarkable technical challenges subject of extensive ongoing research, such as the energy efficient design of such type of antennas, and the so-called two-sided beam alignment problem \cite{yan2019performance,song2020fully}.
The present study assumes that high-gain steerable beamforming antennas are indeed feasible, and that a beam alignment procedure can be successfully performed. Under this assumption, the main focus is the evaluation of the possibly beneficial impact of using highly directive antennas on channel equalization.

\subsection{The impact of directivity on inter-symbol interference}
\begin{figure*}[!t]
\centering
\subfloat[]{\includegraphics[width=.45\linewidth]{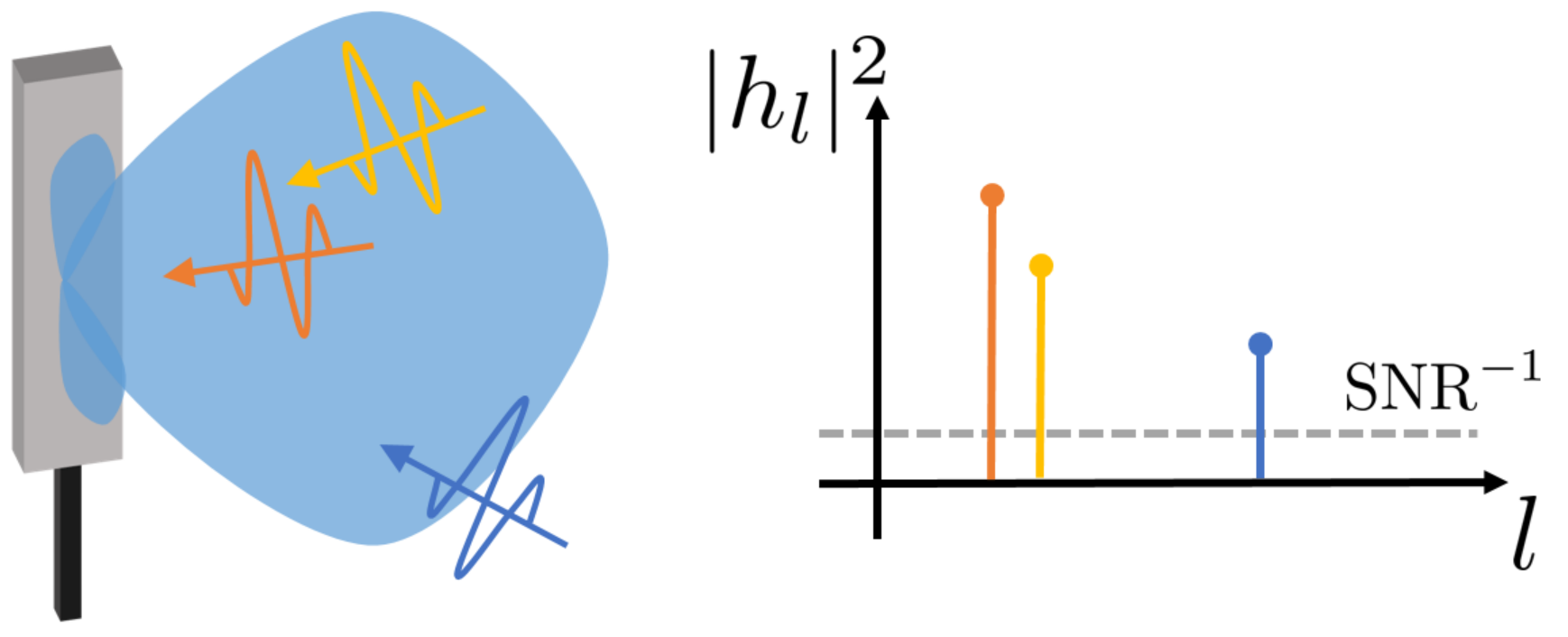}%
\label{fig:wide}}
\hfil
\subfloat[]{\includegraphics[width=.45\linewidth]{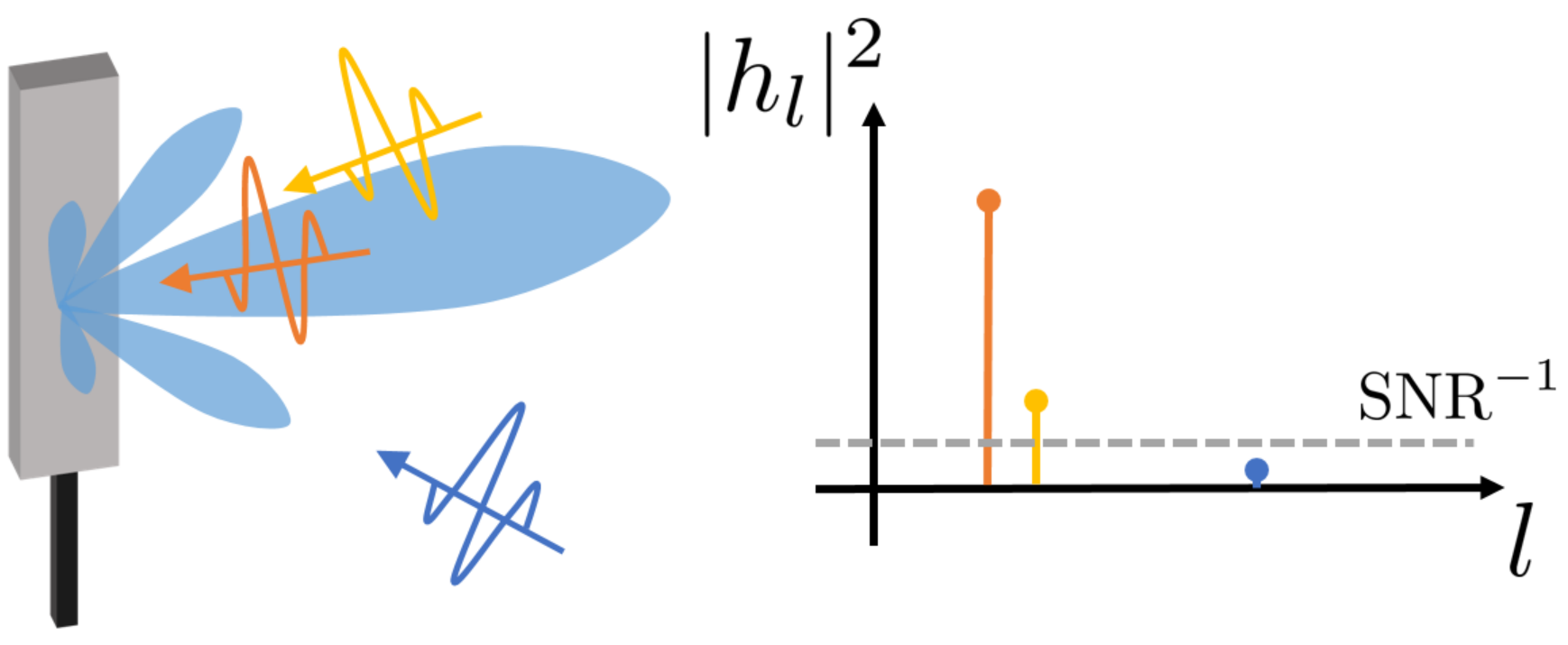}%
\label{fig:narrow}}
\caption{Pictorial representation of the radio channel comprising of a multipath propagation channel and (a) low directivity antennas; (b) high directivity antennas. In contrast to case (a), using highly directive antennas as in case (b) significantly attenuates most multipath components of the propagation channel. As shown experimentally in the remainder of this study, this effect  reduces inter-symbol interference, and hence simplifies equalization.}
\label{fig:spatial_filtering}
\vspace{-0.4cm}
\end{figure*}
Many previous studies on millimiter-wave and THz systems already recognized that the use of highly directive antennas leads to radio channels (i.e., comprising the effect of the antennas) with considerably lower delay spread than the underlying propagation channels, and hence to radio channels for which simple modulation and equalization techniques are potentially sufficient \cite{sun2018millimeter,kurner2008,sarieddeen2021thz}. This crucial observation can be explained by the so-called spatial filtering effect, which is illustrated in Figure~\ref{fig:spatial_filtering}. As customary, we model the single-input single-output radio channel between two beamforming antennas by means of a channel impulse response $h=(h_0,\ldots,h_{L-1})$ with $L$ taps, and a nominal SNR parameter. 
Then, by assuming standard coding techniques, and by treating inter-symbol interference as noise, an achievable spectral efficiency over this channel is
\begin{equation}\label{eq:SE}
\mathrm{SE} = \log_2\left(1+\dfrac{|h_{l_0}|^2}{\sum_{l\neq l_0}|h_l|^2 +\mathrm{SNR}^{-1}} \right),
\end{equation}
where $l_0$ is the delay corresponding to the main peak of the power delay profile. 
Importantly, we remark that the above  transmission scheme does not consider any form of digital equalization. For this reason, in the presence of significant inter-symbol interference, as depicted in Figure~\ref{fig:wide} for the case of low antenna directivity, the spectral efficiency may be quite poor. However, when highly directive antennas are employed and most multipath components are consequently attenuated, as depicted in Figure~\ref{fig:narrow}, this simple scheme may be sufficient to achieve the relatively low target spectral efficiencies of sub-THz mobile access networks. Moreover, if needed, simple equalization techniques may be sufficient to suppress most of the residual inter-symbol interference and further improve upon \eqref{eq:SE}. The reminder of this study is devoted to a deeper investigation of this intuition.

\subsection{The impact of directivity on multi-user interference}
The spectral efficiency expression \eqref{eq:SE} based on treating inter-symbol interference as noise can be readily extended such that additional interference terms corresponding to the incoming signal of other users are included. Following the same spatial filtering argument as for the inter-symbol interference, the use of highly directive antennas may significantly mitigate the impact of multi-user interference, provided that the interfering signal components do not overlap with the main desired signal component. In this work, for simplicity, we neglect multi-user interference by invoking the aforementioned spatial filtering effect, and by assuming the concurrent scheduling of users that are sufficiently separated in the angular domain. Therefore, the resulting capacity figures are only valid up to some maximum multiplexing gain $M$ that, roughly speaking, depends on the antenna beamwidth and the angular distribution of the users.

\subsection{Objective, methodology, and organization of the paper}
Following the above discussion, the main objective of this study is to validate experimentally the claim that the combination of a relatively low target per-stream spectral efficiency and the use of highly directive beamforming antennas may allow the use of simpler modulation and equalization techniques than in traditional designs, and hence leave room for significant energy savings. To this end, differently than in most related studies (see, e.g., \cite{molisch2022directionally,xing2021millimiter}), we follow an unconventional yet more targeted methodology that integrates accurate measurement data with rigorous information theoretic arguments. More specifically, instead of extracting channel modelling parameters such as delay spread, we present the results of a measurement campaign directly in terms of theoretically achievable performance, for different choices of modulation and equalization techniques. The reminder of this study is organized as follows: Section~\ref{sec:measurements} illustrates the performed measurement campaign, and discusses how the obtained measurements can be used to approximate realistic sub-THz communication channels. Section~\ref{sec:eq} presents the results of the measurement campaign in terms of achievable spectral efficiency and coded bit error rate (BER) for promising modulation and equalization techniques. Finally, Section~\ref{sec:conclusion} summarizes the main conclusions and outlines some possible future directions.    

\section{Measurement campaign}\label{sec:measurements}
This section illustrates the technical details of a measurement campaign aiming at characterizing the performance of simple modulation and equalization techniques for realistic indoor radio channels with \SI{160}{\giga\hertz} carrier frequency, \SI{4}{\giga\hertz}  bandwidth, and directive beamforming antennas.

\subsection{Channel sounder setup}
The adopted channel sounder is based on the principle of time-domain channel measurement. Test and measurement equipment working in the millimeter-wave frequency range is expanded with active D-band front-ends that allow extension to carrier frequencies between \SI{110}{\giga\hertz} and \SI{170}{\giga\hertz}. The setup can be divided into transmitter (TX) and receiver (RX), where the TX is kept fixed and the RX is moved across different positions. The TX consists of a signal generator (R{\&}S{\textsuperscript{\textregistered}} SMW200A) and an external D-Band front-end (R{\&}S{\textsuperscript{\textregistered}} FE170ST). The signal generator provides a perfect periodic correlation sequence (Frank-Zadoff-Chu sequence) with a measurement bandwidth of \SI{4}{\giga\hertz} and a sequence duration of \SI{50}{\micro\second} at an intermediate frequency. The external front-end then up-converts the provided signal to the desired carrier frequency of \SI{160}{\giga\hertz}. As transmitting antenna, a standard gain horn with a gain of \SI{25}{\deci\belisotropic} is used. The RX consists of a signal analyzer (R{\&}S{\textsuperscript{\textregistered}} FSW43) and another external D-Band front-end (R{\&}S{\textsuperscript{\textregistered}} FE170SR). The external front-end on the RX down-converts the received signal to an intermediate frequency. The signal analyzer samples the received signal and stores the IQ samples. The synchronization between transmitter and receiver is achieved by using a 
reference signal of \SI{10}{\mega\hertz}. 
The receiver is triggered synchronously to the sequence start by a versatile pulse source (Synchronomat). Hence, coherent sampling and the measurement of time-of-flight are ensured. An open waveguide with \SI{6}{\deci\belisotropic} gain serves the receiving antenna. The chosen antennas emulate a typical sub-THz system with a large steerable antenna at the access point and a smaller steerable antenna at the user equipment after beam alignment.  

\subsection{Measurement scenario and procedure}
\begin{figure}[t]
\begin{center}
  {\includegraphics[width=\columnwidth, trim={0 1.5cm 0 0}, clip]{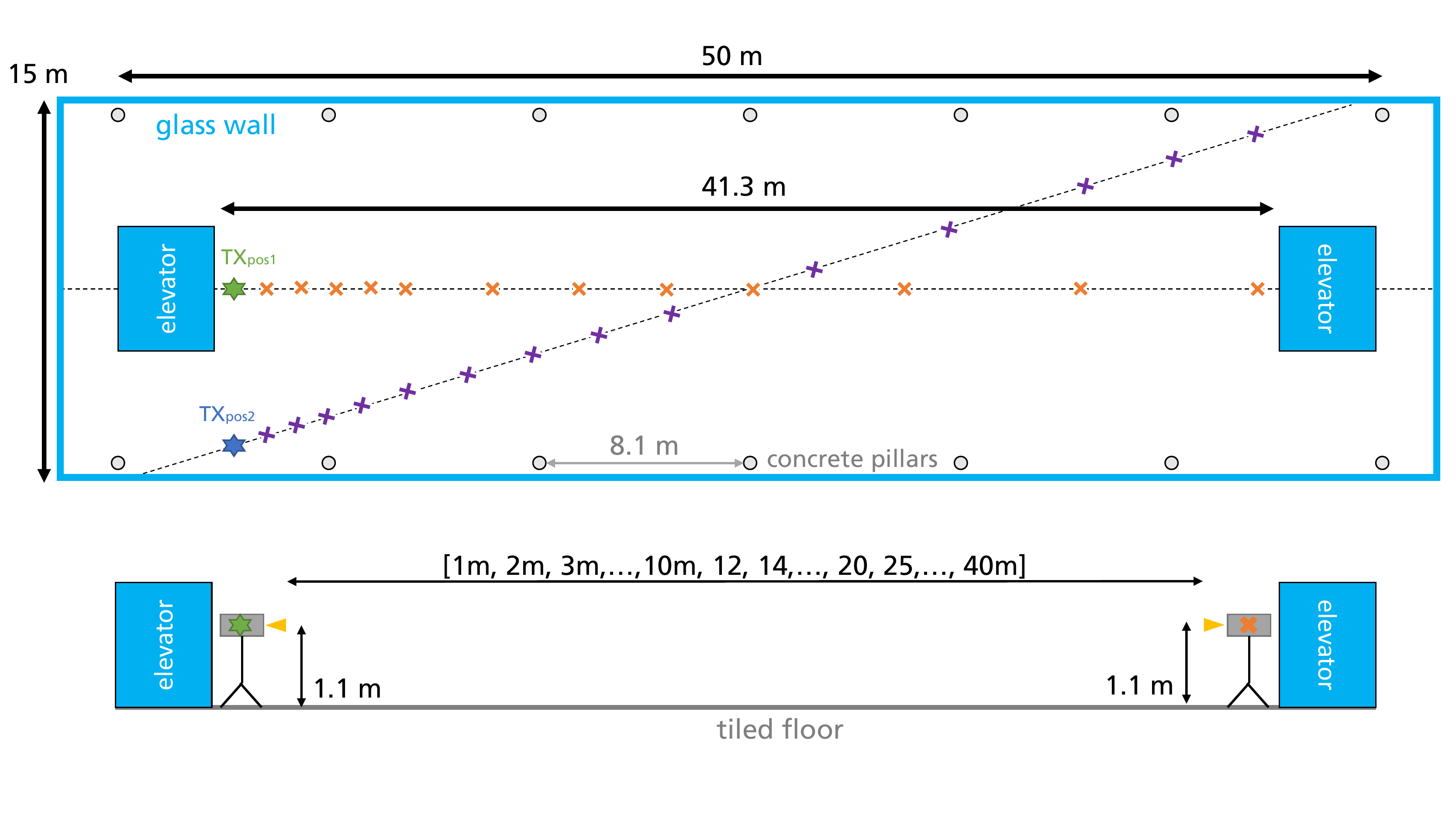}}
  \caption{Schematic visualization of the measurement scenario. The upper drawing is a top-down illustration of the atrium's dimension and composition. The green star represents the TX's first position and the blue star it's second. The dashed lines indicate both the TX's as well the RX's orientation during the measurement campaign's execution. The lower drawing illustrates the TX's and RX's height, as well as the measured distances.}
  \label{fig:floorplan}
\end{center}
\vspace{-0.5cm}
\end{figure}

The measurement campaign is conducted on the company premises of Rohde \& Schwarz in Munich, Germany. As venue, the rectangular atrium of a large building is used. The atrium's dimensions are \SI{15}{\meter} x \SI{50}{\meter} with a height of approximately \SI{20}{\meter}. The environment is mainly characterized by glass walls, a tiled floor, metallic surfaces (elevators) and concrete pillars. 
Figure~\ref{fig:floorplan} illustrates the measurement setup in terms of a floor plan, and the chosen TX and RX positions. For all positions, a line-of-sight (LoS) between TX and RX was ensured, and the TX and RX antennas were aligned along the corresponding LoS direction. This corresponds to the assumed use case where the beams of the steerable antennas at the TX and the RX are aligned. Two different measurement scenarios were considered, differing in the movement pattern within the atrium. In the first scenario, the TX was placed centrally at the beginning of the atrium, and the RX was moved longitudinally. Distances from \SI{1}{\meter} up to \SI{40}{\meter} were considered, resulting in $20$ measurement positions. In the second  scenario, a similar movement pattern was used along the diagonal line connecting two opposite corners of the atrium. Distances from \SI{1}{\meter} to \SI{46}{\meter} were considered, resulting in $26$ measurement positions. 

The first scenario was designed such that multipath components may be detectable, including ground reflections and reflections by, e.g., the metal surfaces of the elevators. The second scenario was designed such that fewer multipath components should be detectable with respect to the first scenario. Nevertheless, the ground reflection may still be detectable.

The collected raw data are processed as follows: First, for each measurement position and scenario, a set of $500$ calibrated channel impulse responses are extracted from the IQ samples, each corresponding to a single sounding sequence. Across the $500$ measurements a phase deviation is estimated under the assumption of a stationary radio channel. Thus, the common phase error of each sequence is compensated and the $500$ complex impulse responses are averaged, resulting in one common impulse response per measurement position with improved correlation gain. Finally, the obtained impulse responses are power normalized, and synchronized with respect to the the first peak. Accurate synchronization with sub-sample resolution is achieved via upsampling and cubic interpolation of the main peak. The resulting channel impulse responses correspond to channels as seen by a hypothetical communication system after phase and frequency offset compensation and time synchronization. Furthermore, since the impulse responses show a dynamic range of more than \SI{60}{\deci\bel} they can be considered as noise free channel estimates for all SNR regions of practical communication systems.

\subsection{Illustrative examples of measured channels}
\begin{figure}[t]
\begin{center}
  {\includegraphics[width=1\columnwidth, trim={550 10 525 25}, clip]{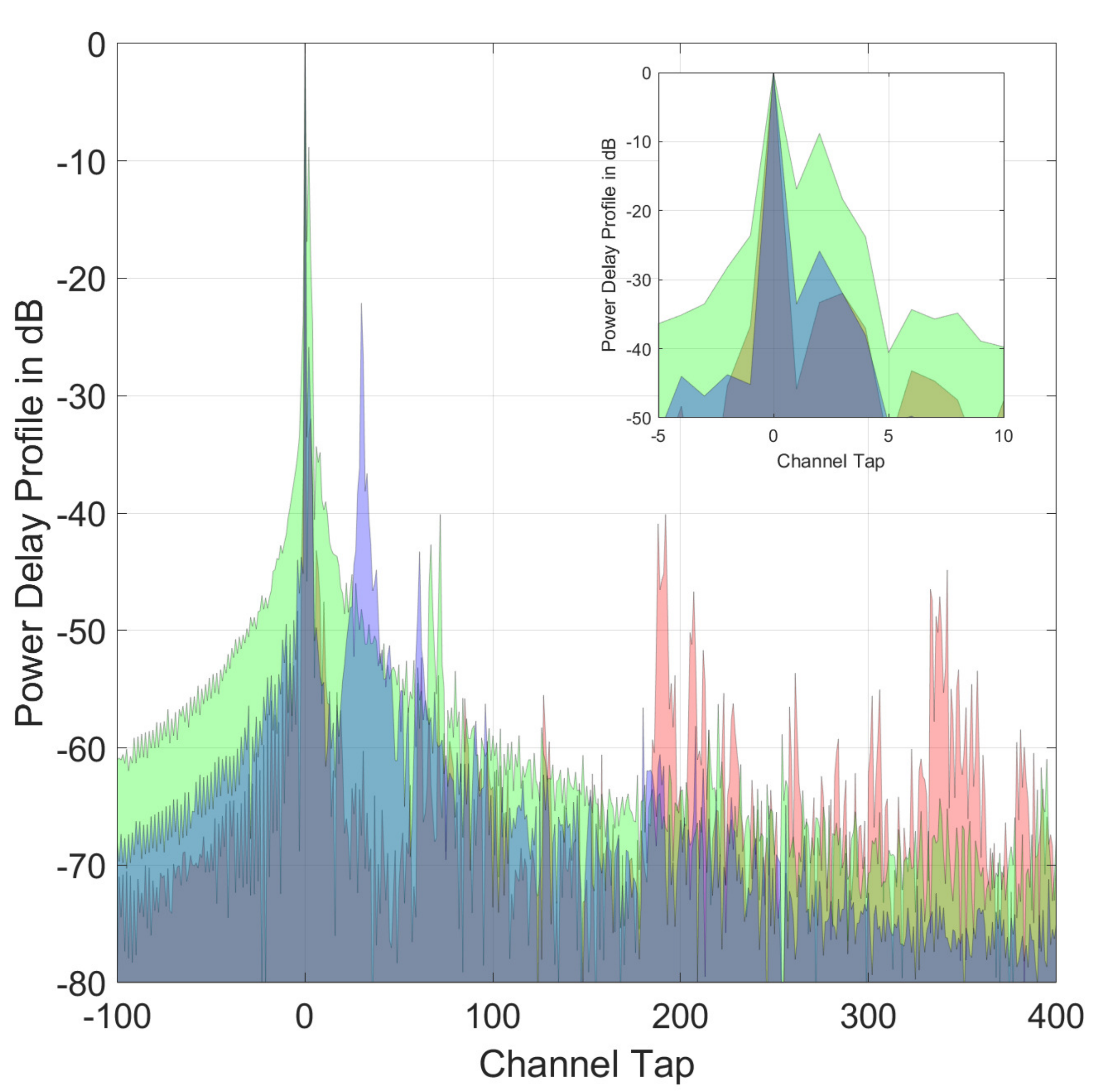}}
  \caption{Power delay profile of three distinguishing measurements (red, green and blue) with enlarged range around the first peak (upper right corner). The time axis is given in channel taps as integers (one tap corresponds to \SI{0.25}{\nano\second}).}
  \label{fig:cir_zoom}
\end{center}
\vspace{-0.5cm}
\end{figure}
The exceptionally high dynamic range  allows the observation of multipath components that are significantly smaller than the LoS path. Additionally, the temporal resolution resulting from the measurement bandwidth of \SI{4}{\giga\hertz} allows the distinction of paths with a theoretical minimum distance of \SI{0.25}{\nano\second}. Figure \ref{fig:cir_zoom} shows the measured power delay profile $|h|^2$ for three representative cases. The green plot refers to a measurement from the first scenario, at the distance of \SI{18}{\meter}. Importantly, the green plot reveals the presence of a strong multipath component only $2$ channel taps away from the LoS path, and it is a representative example of a channel with significant inter-symbol interference. This path is likely due to a ground reflection. The red plot and the blue plot refer to two measurements from the second scenario, at a distance of of \SI{7}{\meter} and \SI{1}{\meter}, respectively. In the red plot, the LoS path is clearly separable from the other paths, which all lie well below \SI{-30}{\decibel}. This is a representative example of a particularly well behaved channel with negligible inter-symbol interference. In the blue plot, two relatively strong multipath components can be observed within a range of \SI{0}{\deci\bel} and \SI{-30}{\deci\bel}, at a delay of $2$ and $30$ taps, respectively. Both components, and especially the second path, may be caused by the interaction with the metal surfaces of the TX and RX devices themselves. This is a representative example of a channel with several significant multipath components with intermediate strengths that might influence the performance of a communication system. Due to the high dynamic range of the measurement setup, more multipath components are visible in all three channels below \SI{-40}{\deci\bel}. Nonetheless, these components are so weak that they can be neglected for the low spectral efficiency regime.

Overall, all measured radio channels appear to be composed only by few relevant propagation paths with significant strength. These propagation paths, corresponding to the peaks of the power delay profiles, are mostly due to reflections captured by the main lobes of the antenna radiation patterns, and are generally located within $10-30$ taps. This confirms the main intuition illustrated in Figure~\ref{fig:spatial_filtering}. However, differently than in the idealized model of Figure~\ref{fig:spatial_filtering}, the channels exhibit additional taps in the close proximity of the channel peaks, which are likely due to the chosen pulse shape.  In fact, the measurements are hard band-limited, i.e., they consider a sinc-like Nyquist pulse with pronounced side lobes, which contribute to inter-symbol interference in the (almost sure) case of the path delay being not perfectly aligned with the sampling grid.

\subsection{Relation with realistic sub-THz communication channels}
As already anticipated, in the reminder of this study we assess the performance of simple modulation and equalization techniques directly over the measured channel impulse responses. A natural question is: to what extend these impulse responses reflect the channel seen by a realistic communication system? First, we point out that the chosen antenna directivities ($25$ dBi and $6$ dBi) can be reasonably implemented using steerable phased arrays of practical size and cost, and, as discussed in Section~\ref{sec:intro}, are almost necessary to provide sufficient coverage at sub-THz frequencies. Note that, since beam alignment is implemented by means of mechanical steering (a form of true-time-delay beamforming), our measurements are insensitive to the spatial wideband effect \cite{wang2018} for the LoS path. Nevertheless, the chosen antenna directivities and signal bandwidth fall into a regime where the spatial wideband effect is negligible for $\lambda/2$ spaced phased arrays, and therefore the measurements are also representative of more practical beamforming implementations. Second, we argue that the chosen pulse shape still produces good approximations of realistic communication channels. In fact, although practical communication systems may consider pulses with faster decay, such as a root raised cosine pulses, the difference would be mostly noticeable for the weaker secondary paths, since accurate time recovery is performed with respect to the main peak. 
Finally, we remark that the current channel impulse responses are representative of a system with $4$ GHz communication bandwidth, which is a reasonable choice for early prototypes of sub-THz mobile access networks. Nevertheless, this bandwidth is already sufficient to capture the main impact of the absolute delay spread with a sub-ns resolution. Therefore, the results in this study could also be approximately extended to systems with higher bandwidths, provided that the delay spread expressed in number of channel taps is scaled accordingly.


\section{Measurement results}\label{sec:eq}
In this section we present the results of our measurement campaign in terms of achievable performance of promising modulation and equalization techniques over the measured channels. In particular, motivated by the importance of energy-efficient designs, we study two alternatives to traditional implementations of multi-carrier modulation based on orthogonal frequency-division multiplexing (OFDM). Multi-carrier modulation is known to approach the fundamental spectral efficiency limits of inter-symbol interference channels with a relatively simple equalization effort, provided that the number of subcarriers is large enough. However, its main drawback is the known issue of producing signals with peak-to-average power ratio (PAPR) proportional to the number of subcarriers, that needs to be compensated by significantly reducing the energy efficiency of the transceiver front-end. Therefore, we consider single-carrier modulation with linear equalization and multi-carrier modulation with a low number of subcarriers. We then show that both alternatives achieve spectral efficiencies of about $1-3$\,bit/s/Hz over all measured channels. Hence, both schemes may be sufficient to score the target capacity gains in a more energy efficient manner.


\begin{figure}[!t]
\centering
\includegraphics[width=1\columnwidth]{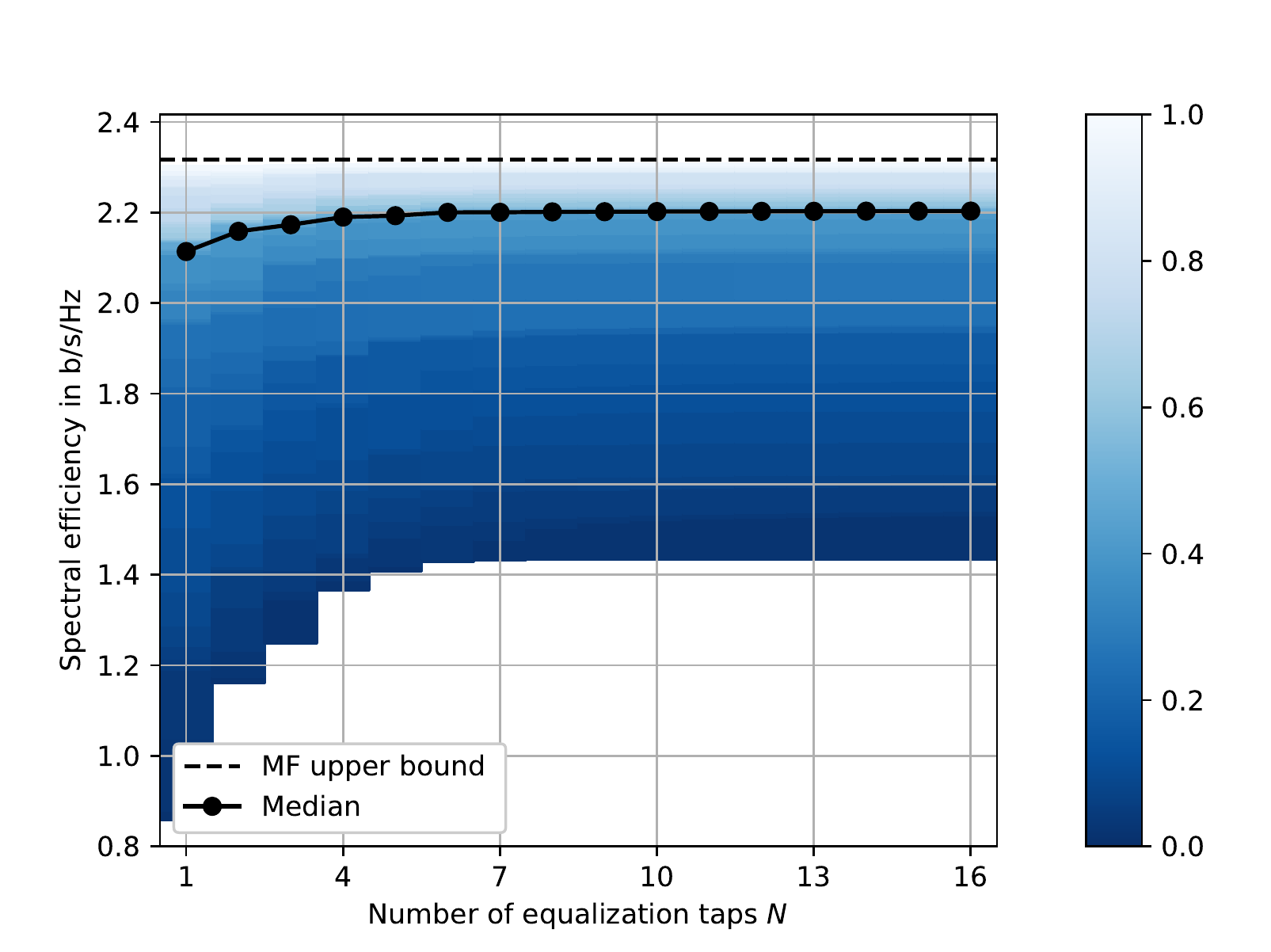}%
\caption{Achievable spectral efficiency for single-carrier modulation with $N$-taps linear equalization, in the considered indoor scenario assuming $\mathrm{SNR}=6$\,dB. The line plots are the median value over all measurement points, and the matched filter upper bound. The variations across measurement points are captured by the empirical cumulative density function, illustrated through a color map, and truncated within the range of observed spectral efficiencies. Due to the specific structure of the measured channels, single-carrier modulation with very few equalization taps appears sufficient to achieve spectral efficiencies of about $1.4-2.3$\,bit/s/Hz, and, hence, score the capacity gains promised by sub-THz networks in an energy efficient manner.}
\label{fig:SC}
\vspace{-0.4cm}
\end{figure}

\subsection{Single-carrier with linear equalization}
We first consider single-carrier modulation followed by a $N$-tap linear equalizer. An achievable spectral efficiency is given by \eqref{eq:SE}, with $h$ replaced by its convolution with the equalizer impulse response. We compute the optimal equalizer coefficients in terms of spectral efficiency, i.e., maximizing the corresponding Rayleigh quotient appearing in \eqref{eq:SE}. The  solution takes the form of a standard linear minimum mean-square error equalizer. We set $l_0$ as the delay of the main peak of the original power delay profile, plus an additional offset which is tuned such that the equalizer focuses on the optimal number of precursors and postcursors (in practice, only one or two postcursors are relevant).

Figure~\ref{fig:SC} reports the resulting achievable spectral efficiency over the measured channels, as a function of the number of taps $N$, where $N=1$ stands for no equalization, and by assuming $\mathrm{SNR}= 6$\,dB. Since the channel impulse responses are normalized, the SNR parameter can be interpreted as a receive SNR comprising of transmit power, path loss, beamforming gain, and receiver noise. The dashed line shows the so-called matched filter upper bound, equivalent to the capacity of an AWGN channel with the same SNR, which provides an upper bound on the achievable spectral efficiency. We remark that this bound is generally not achievable and may be even quite far from the achievable performance under ideal equalization, in channels with significant inter-symbol interference. The chosen SNR is in the high range of what is expected to be practically feasible at the edge of a sub-THz mobile access network. Furthermore, it leads to an upper bound within the spectral efficiency regime of interest, i.e., $1-3$\,b/s/Hz. For lower SNR values, the gains of the equalization decrease until they become negligible. A higher SNR improves performance for all choices of number of taps $N$, up to saturation points driven by the uncompensated inter-symbol interference for the given choice of $N$. 

A we can see, $50\%$ of the measured channels are so well-behaved that no equalization is actually needed to approach almost ideal performance. Furthermore, reliable communication above $1$\,bit/s/Hz without equalization is theoretically possible in the vast majority of the cases. However, in the other $50\%$ of the cases, the inter-symbol interference caused by strong paths in the close vicinity of the main peak causes large variations from the ideal performance. In a few extreme cases the achievable spectral efficiency may even drop below $1$\,bit/s/Hz. Fortunately, this residual inter-symbol interference can be significantly mitigated by using only $5-6$ equalization taps, for which the achievable spectral efficiency is consistently kept above $1.4$\,bit/s/Hz. One important comment here is that, in all measured channels, further spectral efficiency gains can only be observed after adding several tens or even hundreds of equalization taps, i.e., when the filter length reaches the weaker multipath components located far away from the main peak. Nevertheless, these gains are  negligible and do not justify the longer equalizers.   

As a final remark, we recall that the above figures are valid for a $4$\,GHz communication bandwidth, i.e., for theoretical rates on the order of $4$\,Gbit/s/stream. If higher bandwidths are used, and the target SNR is kept to the same value, the number of required equalization taps for each measurement point may increase proportionally, reaching few tens of taps in the worst cases. Still, the main message of this section would continue holding, i.e., single-carrier modulation and a simple linear equalizer may be sufficient to score the target capacity goals in the considered indoor scenario. 

\subsection{Multi-carrier with a low number of subcarriers}
\begin{figure}[!t]
\centering
\includegraphics[width=1\columnwidth]{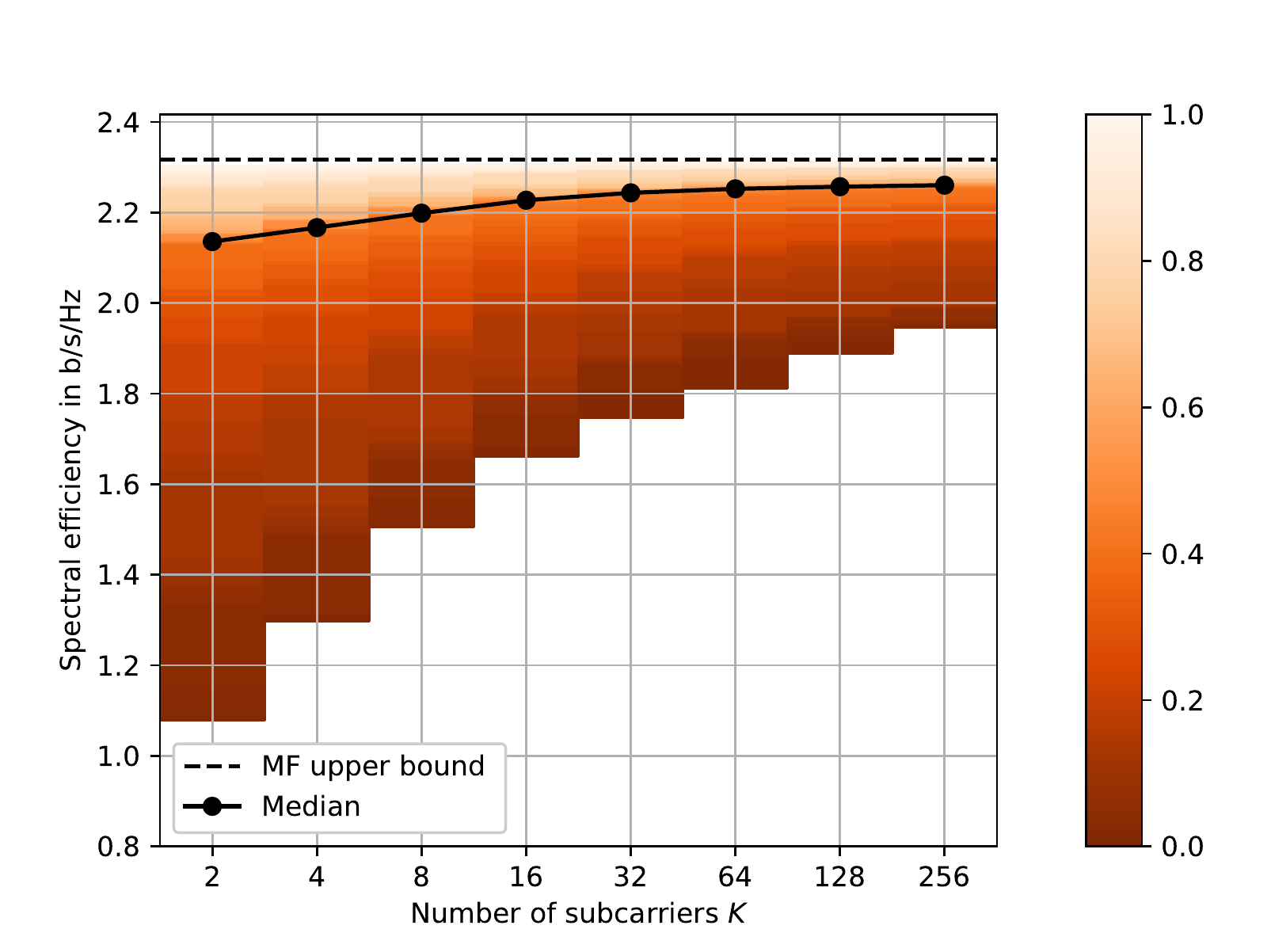}%
\caption{
Achievable spectral efficiency for multi-carrier modulation with $K$ subcarriers and no prefix, in the considered indoor scenario assuming $\mathrm{SNR}=6$ dB. The line plots are the median value over all measurement points, and the matched filter upper bound. The variations across measurement points are captured by the empirical cumulative density function, illustrated through a color map, and truncated within the range of observed spectral efficiencies. Using a low number of subcarriers and no prefix is sufficient to score the target capacity goals, with a much better PAPR than conventional OFDM.}
\label{fig:MC}
\vspace{-0.5cm}
\end{figure}
In this section we consider multi-carrier modulation based on a standard OFDM implementation with $K$ subcarriers and a cyclic prefix of length $Q$. In contrast to traditional designs, we study the effect of choosing a prefix length much shorter than the delay spread, i.e., we focus on $Q\ll L$. In this regime, the channel in the subcarrier domain suffers from both inter-carrier interference (ICI) and inter-block interference (IBI) \cite{molish2000ofdm}. By following similar arguments as for the single-carrier case, i.e., by treating ICI and IBI as noise, an achievable spectral efficiency is given by the sum of $K$ expressions similar to \eqref{eq:SE}, where the signal and inter-symbol interference terms are replaced by the appropriate channel coefficients in the subcarrier domain, and a pre-log factor which takes into account the prefix overhead. For all tested numbers of subcarriers $K$, we optimize the prefix length $Q$. Furthermore, similarly to the single-carrier case, we consider a block-boundary detection stage such that the prefix is optimally centered with respect to the main channel peak. 

Interestingly, for all tested numbers of subcarriers, the gains of adding a cyclic prefix in terms of ICI and IBI mitigation are minor and outweighed by the multiplicative spectral efficiency losses due to the overhead. As a consequence, we focus on an unconventional implementation with no prefix ($Q=0$), as proposed in \cite{molish2000ofdm}. However, differently than \cite{molish2000ofdm}, we do not consider additional equalization stages, and simply treat ICI and IBI as noise. Similar to the single-carrier case, Figure~\ref{fig:MC} reports the resulting achievable spectral efficiency over the measured channels as a function of the number of subcarriers $K$ and by assuming $\mathrm{SNR}= 6$\,dB. We notice that very few subcarriers are needed to achieve spectral efficiencies significantly higher than $1$\,bit/s/Hz in all measured channels and are often sufficient to be close to the ideal performance limit. An informal explanation is that, even without a cyclic prefix, the inter-symbol interference is spread over multiple subcarriers and pushed towards the noise floor as $K$ increases. Furthermore, at the price of a higher PAPR and hence of more stringent hardware efficiency requirements, we observe that the proposed multi-carrier modulation for $K\geq 16$ achieves better spectral efficiencies (in both median and worst case) than single-carrier modulation even for a large number of equalization taps (we tested up to $N=256$ equalization taps). 

\subsection{Performance of a complete system}
\begin{figure}[!t]
\centering
\includegraphics[scale=0.5]{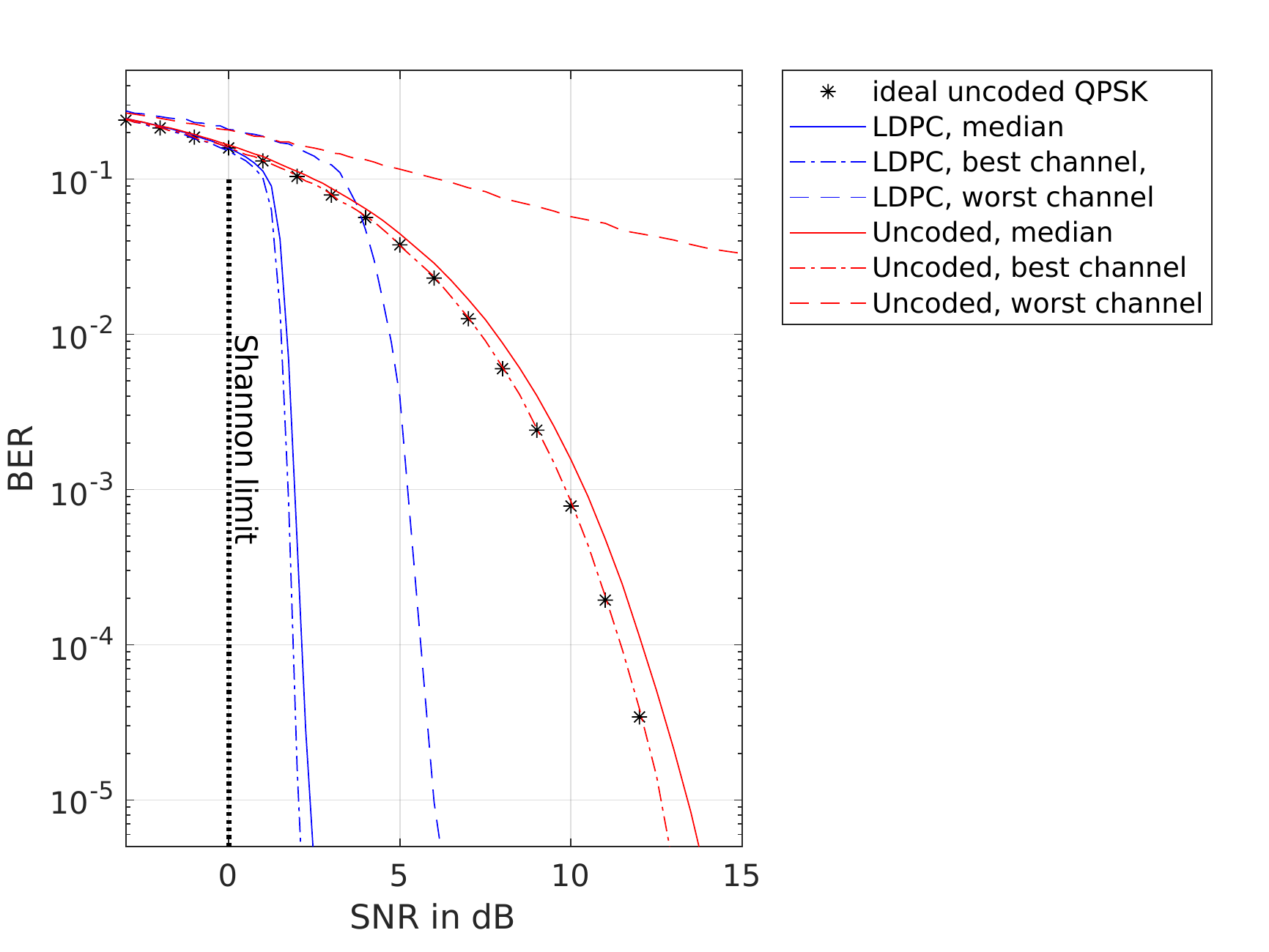}%
\caption{BER versus SNR for single-carrier QPSK modulated signals and $7$-taps linear equalization, in the considered indoor scenario. The median, worst, and best performance over all measured points are shown. The coded BER curves refer to standard LDPC codes with rate 1/2 and block length $1296$, and show that reliable communication at $1$ bit/s/Hz is indeed feasible by means of simple and energy-efficient modulation and equalization techniques.}
\label{fig:ber}
\vspace{-0.5cm}
\end{figure}
In this section, we show that reliable communication at \SI{1}{bit/s/Hz} over the measured channels is indeed feasible by using simple modulation and equalization techniques. Specifically, we run a numerical simulation of a system employing single-carrier QPSK modulated signals, linear minimum mean-square error equalization with $N=7$ taps (two of which are postcursors), and standard low-density parity-check (LDPC) codes. The LDPC code has a rate of $1/2$, a $1296$\,bit block length, and uses the parity-check matrix as defined in the IEEE 802.11n standard. The LDPC decoder is tuned by treating inter-symbol interference as additive Gaussian noise with the same power and uses a belief propagation algorithm. The simulated transmit signals are distorted by the measured channel impulse responses and Gaussian noise is added according to the SNR parameter. Figure~\ref{fig:ber} reports the performance in terms of coded BER for all measured channels as a function of the SNR. For reference, we also report the corresponding uncoded BER curves, and a lower bound on the minimum SNR for reliable communication at $1$\,b/s/Hz, obtained from the aforementioned matched filter upper bound on the achievable spectral efficiency (Shannon limit). Remarkably, the gap between the obtained BER curves and the (not necessarily achievable) Shannon limit is in the vast majority of cases between $2.5-4$ dB, a range which is in line with the typical gap from capacity of LDPC codes on AWGN channels. Furthermore, letting $\mathrm{SNR}\approx 6$ dB allows for reliable communication even for the worst-case channels, as predicted by Figure~\ref{fig:SC}. This confirms that, for low target spectral efficiencies, inter-symbol interference can be indeed successfully mitigated by means of simple equalization techniques, at the point that it can be treated as noise and further mitigated by standard  codes for the AWGN channel.

\section{Conclusion}\label{sec:conclusion}
Our results corroborate the attractiveness of simple and energy efficient alternatives to traditional OFDM-based designs in sub-THz mobile access networks. In particular, our results strengthen the claim that single-carrier modulation is a promising solution for the early development of these networks, where energy efficiency is a major limiting factor. Then, our results suggest that an evolution towards more capable networks can be accomplished by gradually increasing the number of subcarriers in an unconventional OFDM implementation. Our methodology integrates accurate experimental data with rigorous information theoretic arguments. The integration of these approaches is particularly effective, since our channel measurements closely approximate the impulse response as seen by a hypothetical yet realistic sub-THz communication system in the same indoor environment, and with the same signal bandwidth ($4$\,GHz), carrier frequency ($160$\,GHz), and antenna directivities at the access point ($25$\,dBi) and user equipments ($6$\,dBi). To further test the robustness of our analysis, future works should explore different system parameters. For instance, even though some preliminary conclusions could already be extrapolated from this study, a direct performance evaluation for larger bandwidths is desirable. Perhaps most importantly, future works should also cover different environments with more reflections and possibly a higher number of multipath components. Furthermore, non-line-of-sight propagation and the effect of multi-user interference are interesting additions to the analysis. 

\bibliographystyle{IEEEbib}
\bibliography{IEEEabrv,refs}
%

\section*{Biographies}
%
%
%

\noindent\footnotesize \textbf{Lorenzo Miretti} received the B.Sc. and M.Sc. degrees in Telecommunication Engineering from Politecnico di Torino in 2015 and 2018, respectively, and the Ph.D. degree in wireless communications from EURECOM and Sorbonne Université in 2021. He is currently a post-doctoral researcher with the Technical University of Berlin and the Fraunhofer Heinrich Hertz Institute. He investigates novel solutions for next generation wireless networks, such as cell-free massive MIMO and sub-THz mobile access networks.
\vspace{0.3cm}

\noindent\footnotesize \textbf{Thomas Kühne} received the Dipl.-Ing. (equivalent to an M.Sc.) degree in Electrical Engineering from the University of Technology Dresden, Germany, and the Dr. -Ing. (Ph.D.) degree in 2022 in wireless communication from the Technische Universität Berlin, Germany. During his master's studies, he focused on communication systems and circuit design, while during his Ph.D. studies, he specialized in wireless communication systems and signal processing. Since 2015, he has been working in Prof. Caire's Communication and Information Theory group at Technische Universität Berlin, currently as a Postdoc researcher. His research interests include the co-design between the hardware, the signal processing, and the system design of next generation wireless communication systems.
\vspace{0.3cm}

\noindent\footnotesize \textbf{Alper Schultze} received the B.Sc. degree and the M.Sc. degree in electrical engineering from the Technical University of Berlin, Germany, in 2016 and 2019, respectively. During his studies, he specialized on communication systems and spent one year abroad in Italy at the University of Bologna deepening his knowledge on telecommunication engineering. In 2019, he joined the Fraunhofer Heinrich Hertz Institute in Berlin, as a research associate. His research focuses on channel measurement and characterization in the sub-THz and THz domain.
\vspace{0.3cm}

\noindent\footnotesize \textbf{Wilhelm Keusgen}
received the Dipl.-Ing. (M.S.E.E.) and Dr.-Ing. (Ph.D.E.E.) degrees from the RWTH Aachen University, Aachen, Germany, in 1999 and 2005, respectively.
From 1999 to 2004, he was with the Institute of High Frequency Technology, RWTH Aachen University, where he worked on the reciprocity of multiple antenna systems.
From 2004 to 2021 he was heading a research group for millimeter-waves and advanced transceiver technologies at the Fraunhofer Institute for Telecommunications, Heinrich Hertz Institute, located in Berlin, Germany.
Since 2021 he is a full professor at Technische Universität Berlin and heading the chair for Microwave Systems. His main research areas are millimeter wave and THz communications for 5G and beyond. 
\vspace{0.3cm}

\noindent\footnotesize \textbf{Giuseppe Caire} received his B.Sc. degree from Politecnico di Torino, Italy, in 1990, his M.Sc. degree from Princeton University, New Jersey, in 1992, both in electrical engineering, and his Ph.D. degree from Politecnico di Torino in 1994. He has been a post-doctoral research fellow with the European Space Agency (ESTEC, Noordwijk, The Netherlands) in 1994-1995, Assistant Professor in Telecommunications at the Politecnico di Torino, Associate Professor at the University of Parma, Italy, Professor with the Department of Mobile Communications at the Eurecom Institute,  Sophia-Antipolis, France, a Professor of Electrical Engineering with the Viterbi School of Engineering, University of Southern California, Los Angeles, and he is currently an Alexander von Humboldt Professor with at the Technische Universität Berlin, Germany. Among others, he received an ERC Advanced Grant in 2018, the IEEE Communications Society Edwin Howard Armstrong Achievement Award in 2020, and the 2021 Leibniz Prize of the DFG.
\vspace{0.3cm}

\noindent\footnotesize \textbf{Michael Peter} received the Dipl.-Ing. degree (M.S.) in electrical engineering and information technology from the University of Karlsruhe, Germany, in 2004, and the Dr.-Ing (Ph.D.) degree in electrical engineering, telecommunications, from the Technische Universität Berlin, Germany, in 2018. 
He joined the Fraunhofer Institute for Telecommunications, Heinrich Hertz Institute, Berlin, Germany in 2005. In 2019, he became co-leader of the Millimeter Wave Research Group and took over as head in 2021. His research interests include millimeter-wave and THz communications with focus on channel measurements and modeling, physical layer design and simulation, and performance analysis taking into account hardware impairments. He is author and co-author of more than 70 peer-reviewed scientific papers.
\vspace{0.3cm}

\noindent\footnotesize \textbf{Sławomir~Sta\'nczak} is Professor of Network Information Theory at the Technical University of Berlin and Head of the Wireless Communications and Networks Department at the Fraunhofer Heinrich Hertz Institute (HHI). Prof. Sta\'nczak is co-author of two books and more than 200 peer-reviewed journal articles and conference papers in the field of information theory, wireless communications, signal processing, and machine learning. Prof. Sta\'nczak received research grants from the German Research Foundation and the Best Paper Award from the German Society for Telecommunications in 2014. He was an associate editor of the IEEE Transactions on Signal Processing from 2012 to 2015 and chair of the ITU-T Focus Group on Machine Learning for Future Networks including 5G from 2017 to 2020. Since 2020 Prof. Sta\'nczak is chairman of the 5G Berlin association and since 2021 he is coordinator of the projects 6G-RIC (Research \& Innovation Cluster) and CampusOS.
\vspace{0.3cm}

\noindent\footnotesize \textbf{Taro Eichler} is Technology Manager for wireless communications and photonics at Rohde \& Schwarz in Munich with a focus on 5G/6G technologies. Prior to joining Rohde \& Schwarz, Taro worked for Intel Corporation as specialist for photonics communication solutions and for NTT Basic Research Laboratories. He also worked on research projects at The University of Tokyo in the field of quantum optics. He holds a diploma in physics from the Technical University of Munich with thesis at the Max-Planck-Institute for Quantum Optics and a PhD in physics from the University of Bonn, Germany.

\end{document}